\documentclass[aps,prl,twocolumn,superscriptaddress]{revtex4-1}
\usepackage{graphicx}
\usepackage{dcolumn}
\usepackage{bm}

\begin{document}


\title{Maximizing Fermi surface multiplicity optimizes
superconductivity in iron pnictides}


\author{Hidetomo Usui}
\affiliation{Department of Applied Physics and Chemistry,
The University of Electro-Communication, Chofu, Tokyo 182-8585, Japan}

\author{Kazuhiko Kuroki}
\affiliation{Department of Applied Physics and Chemistry,
The University of Electro-Communication, Chofu, Tokyo 182-8585, Japan}
\affiliation{JST, TRIP, Chiyoda, Tokyo 102-0075, Japan}

\date{\today}

\begin{abstract}
We study the condition for optimizing 
superconductivity in the iron pnictides from the lattice structure 
point of view. Studying the band structure of the 
hypothetical lattice structure of LaFeAsO, 
the hole Fermi surface multiplicity is found to be 
maximized around the Fe-As-Fe bond angle regime where the arsenic atoms form a 
regular tetrahedron. Superconductivity is optimized within this 
three hole Fermi surface regime, while the stoner factor of the 
antiferromagnetism 
has an overall tendency of increasing upon decreasing the bond angle.
Combining also the effect of the varying the Fe-As bond length, 
we provide a guiding principle for obtaining high $T_c$.
\end{abstract}

\pacs{74.62.Bf, 74.20.-z, 74.70.Xa}

\maketitle

The discovery of superconductivity 
in the iron pnictides\cite{kamihara_LaFeAsO} 
and its $T_c$ up to 55K\cite{Ren} 
has given great impact to the field of condensed matter physics.
To attain even higher $T_c$, 
the investigation on the condition for 
optimizing the superconductivity is highly desired.
In this context, much attention has been paid to 
the correlation between $T_c$ and the lattice structure 
from the early stage.
In particular, Lee $\it{et \ al.}$ have shown that $T_c$ systematically 
varies with the Fe-Pn-Fe (Pn=pnictogen) bond 
angle,  and takes its maximum around 109 degrees, at which the pnictogen 
atoms form a regular tetrahedron (``Lee's plot'')\cite{Lee}.
Theoretically, we have previously explained this lattice structure effect 
within a spin fluctuation mediated pairing theory on 
a five orbital model, and pointed out that 
superconductivity is strongly affected by the Fermi surface around
the wave vector ($\pi$,$\pi$) in the unfolded Brillouin zone\cite{Kuroki_prb}.
This Fermi surface has been found to be 
controlled by the pnictogen 
height $h_{Pn}$ measured from the iron plane. 
When the pnictogen is at high positions, a hole Fermi surface
originating 
from the $X^2-Y^2$ orbital\cite{XYZ} appears around ($\pi$,$\pi$), 
and the spin fluctuation arising from the interaction between 
electron and hole Fermi surfaces gives rise to a high $T_c$  
s$\pm$-wave paring, where the gap is fully open but 
changes sign between electron and 
hole Fermi surfaces as was first proposed in ref.\cite{Mazin}.
When the pnictogen position is low, the $X^2-Y^2$ band sinks below the 
Fermi level, and a 
low $T_c$ nodal $s\pm$-wave paring\cite{Graser} or 
$d$-wave paring takes place\cite{Kuroki_prl}.
Thus, the appearance of the $X^2-Y^2$ Fermi surface around 
$(\pi,\pi)$ (which we call 
$\gamma$) is favorable for superconductivity, so that the higher 
pnictogen height results in a higher $T_c$ within this scenario.
In ref.\onlinecite{Kuroki_prb}, 
the saturation of $T_c$ in LnFeAsO in the bond angle regime of
less than 109 degrees\cite{Lee,Miyazawa} 
has been attributed to the decrease of the 
lattice constant $a$, which is found to be unfavorable for 
superconductivity due to the decrease of the density of states
\cite{Kuroki_prb}. 

However, it is becoming clearer that materials having very large $h_{Pn}$ 
of $>1.5\AA$ do not have high $T_c$\cite{Takano}.
Recently, for example, 
Ca$_4$Al$_2$O$_6$Fe$_2$As$_2$ and Ca$_4$Al$_2$O$_6$Fe$_2$P$_2$, 
which are variations of the materials having perovskite 
block layers  (21311 systems)\cite{Ogino}, 
have been found to exhibit superconductivity of 28.3K and 17.1K,
respectively\cite{Shirage}. These materials have very large $h_{\rm Pn}$ 
($\simeq 1.5 \AA$),
small lattice constant $a$ ($\simeq 3.71\AA$), 
and consequently a very small bond angle ($\sim 102$ degrees).
Such a high pnictogen position should give a robust $\gamma$ Fermi
surface around $(\pi,\pi)$ and thus strong spin fluctuations, 
and it seems difficult to understand the 
low $T_c$ even if the effect of the 
small lattice constant is taken into account.
In this context, a recent band structure calculation of 
Ca$_4$Al$_2$O$_6$Fe$_2$As$_2$ have found an interesting feature; 
one of the hole Fermi surfaces around (0,0) is missing, 
resulting in two hole Fermi surfaces\cite{Miyake}. 
This is found to be due to the very small bond angle.

Given this background, in the present Letter, we 
study the condition for optimizing superconductivity 
in the iron pnictides, 
varying hypothetically the lattice structure of 
LaFeAsO. In varying the bond angle $\alpha$ in a wide range while fixing 
the bond length, the number of hole Fermi surfaces 
changes from two to three as $\alpha$ is decreased
(as discussed as the effect of the pnictogen height in previous 
studies), but as $\alpha$ is decreased even further, 
one of the hole Fermi surfaces around $(0,0)$ in the unfolded 
Brillouin zone disappears just as in Ca$_4$Al$_2$O$_6$Fe$_2$As$_2$.
Consequently, the number of hole Fermi surfaces is maximized 
around the bond angle where the arsenic atoms form a
regular tetrahedron.
We note that 
the disappearance of the hole Fermi surface  
was not considered in our previous study\cite{Kuroki_prb} 
since such a small bond angle regime was not considered there.
Applying fluctuation exchange (FLEX) method\cite{Bickers} 
to the model, we find that the superconductivity is optimized in the 
bond angle regime where there are three hole Fermi surfaces. 
Interestingly, the tendency toward magnetism (the strength of the 
spin fluctuation at zero energy) is not necessarily correlated  
with the hole Fermi surface multiplicity, and therefore the correlation 
between superconductivity and the strength of the spin fluctuations 
is complicated.
We also vary the Fe-As bond length while fixing the bond angle,
and find that $T_c$ and the bond length 
is positively correlated mainly due to the increase of the density of 
states without affecting the Fermi surface.

\begin{figure}
\includegraphics[width=7cm]{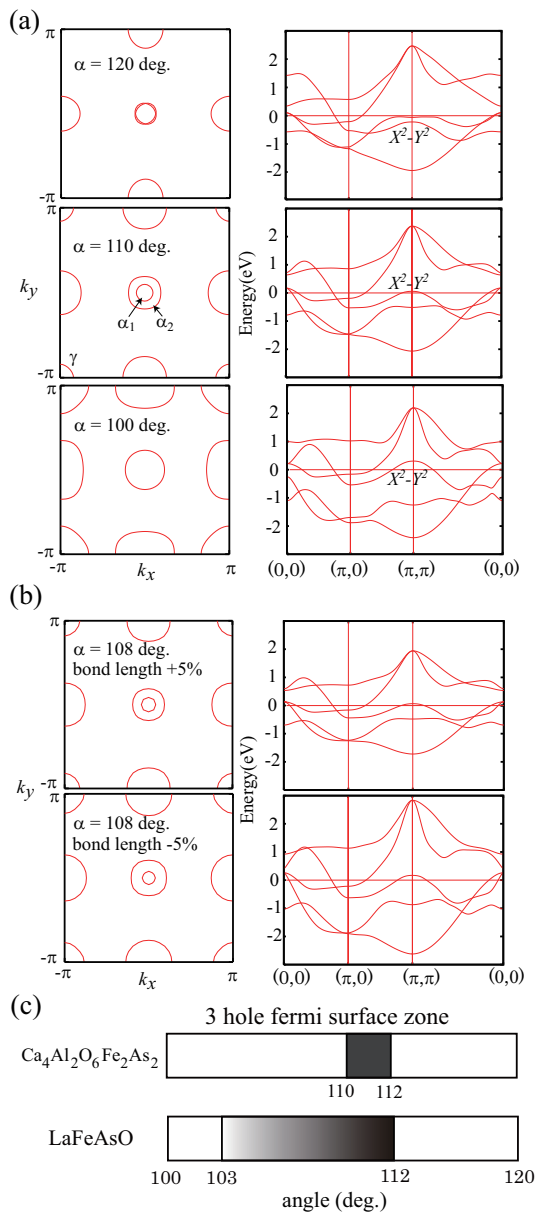}
\caption{\label{fig:1} The band structure of hypothetical lattice 
structures of LaFeAsO with (a) varying the Fe-As-Fe bond angle 
while fixing the Fe-As bond length and (b) vice versa. 
(c) The bond angle range where three hole Fermi surfaces 
coexist is shown for LaFeAsO and Ca$_4$Al$_2$O$_6$Fe$_2$As$_2$.
 }
\end{figure}

We first calculate the band structure of 
hypothetical lattice structures of LaFeAsO using 
the Quantum-ESPRESSO package\cite{pwscf}, and 
construct five band models\cite{Kuroki_prl} 
from maximally localized 
Wannier functions\cite{MaxLoc},
where 
we fix the bond length at its original length\cite{kamihara_LaFeAsO} 
and vary the bond angle $\alpha$(Fig.\ref{fig:1}(a)) as 
has been done for Ca$_4$Al$_2$O$_6$Fe$_2$As$_2$ 
by Miyake {\it et al}\cite{Miyake}.
The Fermi surface is obtained for the band filling(=number of electrons
per site) of $n=6.1$. 
When the bond angle is large, two hole Fermi surfaces, 
$\alpha_1$  and $\alpha_2$ 
are present around the wave vector (0,0). As $\alpha$ decreases, 
the $\gamma$ Fermi surface appears around $(\pi,\pi)$, and we now 
have three hole Fermi surfaces. This is what has
been noticed as an effect of increasing the pnictogen height
\cite{Singh,Vildosola,Kuroki_prb}.
As we decrease $\alpha$ even further, the $\alpha_1$ Fermi surface 
disappears, and again there are only two hole Fermi surfaces but in this 
case one around $(0,0)$ and another around $(\pi,\pi)$.
So the tendency found in Ca$_4$Al$_2$O$_6$Fe$_2$As$_2$ holds also for 
LaFeAsO. However, the range of the bond angle in which there are three hole 
Fermi surfaces (three hole Fermi surface zone in Fig.\ref{fig:1}(c)) 
turns out to be 
much larger for LaFeAsO than for Ca$_4$Al$_2$O$_6$Fe$_2$As$_2$.
We find that the three hole Fermi surface zone tends to be 
small for materials with large $c$-axis length.
Another point that should be mentioned is that the 
total band width is almost unchanged as the bond angle is varied.

Having understood the effect of the bond angle on the 
band structure, we 
next fix $\alpha=108$  and vary the 
bond length (Fig.\ref{fig:1}(b)). 
The band width is reduced upon increasing the 
bond length as expected, but interestingly, 
the Fermi surface is barely affected.
In fact, we find that multiplying the bands for different bond length 
 by certain factors  gives nearly the identical  band structures, 
meaning only the band width, but not the shape, is affected by the bond length.

We now move on to the FLEX calculation. We consider
the standard multiorbital 
interactions (intraorbital and interorbital repulsion, 
Hund's coupling, and the pair hopping interaction),  
and apply FLEX.
In FLEX, bubble and ladder type diagrams consisting of renormalized 
Green's functions are summed up to 
obtain the susceptibilities, which are used to calculate the 
self energy. The renormalized Green's functions are then determined
 self-consistently from the Dyson's equation.
The obtained Green's 
function is plugged into the linearized Eliashberg equation, 
whose eigenvalue $\lambda$ reaches unity at the superconducting transition 
temperature $T=T_c$.  Also, in order to investigate the correlation between 
superconductivity and magnetism, we obtain the stoner factor $a_S$
of the antiferromagnetism 
at the wave vector $(\pi,0)$ in the unfolded Brillouin zone, 
which is defined as the largest eigenvalue of the matrix 
$U\chi_0({\bf{k}}=(\pi,0),i\omega_n=0)$, 
where $U$ is the interaction and $\chi_0$ is the irreducible 
susceptibility matrices, respectively. This value monitors the 
tendency towards stripe type antiferromagnetism and the 
strength of the spin fluctuations at zero energy.
Since the three dimensionality is not strong in LaFeAsO, 
we take a two dimensional model where we neglect the out-of-plane 
hopping integrals, and take $32 \times 32$ $k$-point meshes and  
4096 Matsubara frequencies.
\begin{figure}
\includegraphics[width=7cm]{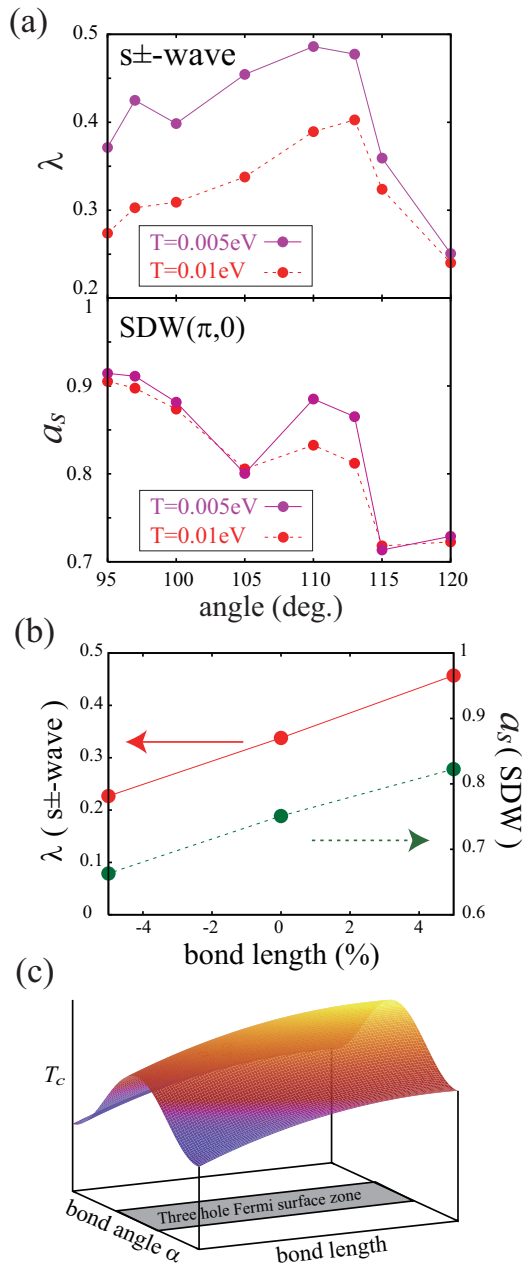}
\caption{\label{fig:2} (a) (Top) the Eliashberg equation eigenvalue 
for s$\pm$-wave pairing and (bottom) the stoner factor at ($\pi$,0) 
against the bond angle for temperatures $T=0.005$ (solid) and 
0.01eV (dashed). 
The interaction reduction factor is $f=0.6$.
(b) The eigenvalue of the Eliashberg equation and the stoner factor 
against the bond length. $T=0.005$eV and $f=0.5$.
(c) A schematic $T_c$ diagram in the bond angle and length space.
}
\end{figure}

As for the electron-electron interaction values, 
we adopt the orbital-dependent interactions
as obtained from first principles calculation 
in ref.\cite{Miyake2} for LaFeAsO, but multiply all of them 
by a constant reducing 
factor $f$. The reason for introducing this factor is as follows.
As has been studied in refs.\onlinecite{Ikeda_prb,Arita,Ikeda_jpsj} 
the FLEX for models obtained from LDA calculations tend to 
overestimate the effect of the 
self-energy because LDA already partially takes into account the 
effect of the self-energy in the exchange-correlation functional. 
When the electron-electron interactions as large as those obtained 
from first principles are adopted in the FLEX calculation, 
this double counting of the self-energy becomes so large 
that the band structure largely differs from its original one. 
In such a case, the spin fluctuations will develop around the wave vector
$(\pi,\pi)$ rather than $(\pi,0)$ 
in the unfolded Brillouin zone, which is in disagreement 
with the experiments. 
In the present study, we therefore 
introduce the factor $f$ so as to reduce the electron-electron interactions.
We vary $f$ within the range where 
the spin fluctuations develop around $(\pi,0)$ and not $(\pi,\pi)$. 
This approach would have problems in accuracy if we try to estimate the 
absolute values of $T_c$ (the approach should underestimate $T_c$), 
but instead here we calculate the eigenvalue of the 
Eliashberg equation and the stoner factor 
at a fixed temperature, thereby comparing the 
{\it relative} strength towards superconducting/magnetic instability 
among various lattice structures. 
%
%

\begin{figure}
\includegraphics[width=7cm]{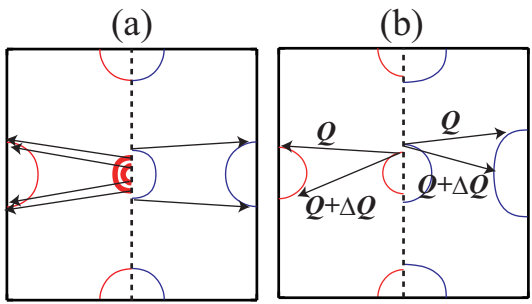}
\caption{\label{fig:3} A schematic figure of (a) the difference 
in the number of pair scattering channels between the cases with 
two small or one large hole Fermi surface(s) (b) the difference 
between small and large Fermi surfaces regarding  
the pair scattering processes mediated by spin fluctuations 
with a momentum spread of $\Delta Q$.
}
\end{figure}

In Fig.\ref{fig:2}(a), we plot the eigenvalue of the 
Eliashberg equation $\lambda$ for the s$\pm$-wave superconductivity 
and the  antiferromagnetic stoner factor $a_S$ at 
($\pi$,0) as functions of the bond angle while fixing the bond length 
at its original value. 
In the large angle regime, the eigenvalue of the Eliashberg equation 
remains relatively small.
As the angle is decreased, the $\gamma$ Fermi surface around $(\pi,\pi)$ 
becomes effective and the eigenvalue becomes large.
This enhancement of superconductivity is correlated with the 
presence/absence of 
nodes in the superconducting gap on the electron Fermi surface (not shown)
as has been studied previously\cite{Kuroki_prb,Graser,Thomale,DHLee}.
As the angle is decreased further, superconductivity tends to be 
suppressed, although the variation of the eigenvalue is not monotonic.
Thus, the superconductivity is optimized 
around 110 degrees, 
which is in agreement with the Lee's plot\cite{Lee}.
The fact that superconductivity is optimized for maximum Fermi 
surface multiplicity can be 
considered as natural since larger number of Fermi surfaces 
gives rise to larger number of pair scattering channels as shown 
in Fig.\ref{fig:3}(a). 
We stress here that due to the quasi-two-dimensionality, 
the density of states is not affected by the size of the Fermi
surface, so two small Fermi surfaces is more favorable than one large 
Fermi surface regarding the number of pair scattering channels. 

On the other hand, the situation is not so simple 
since the stoner factor at $(\pi,0)$,  
which measures the spin fluctuations and thus should also affect $T_c$, 
is found to be 
not necessarily correlated with the multiplicity of the Fermi surface.
In the large angle regime of $115<\alpha<120$, 
superconductivity begins to grow as 
the bond angle is decreased, but the change in the stoner factor is small.
This can be interpreted as follows. 
When the $X^2-Y^2$ band around $(\pi,\pi)$ sits somewhat 
away from he Fermi level, the interaction between this band and the 
electron Fermi surfaces gives rise to spin fluctuations with finite energy 
that contributes to the pairing interaction but has small contribution to 
spin fluctuations with nearly zero energy. 
This can be related to the NMR experiments in LaFeAsO, 
which do not find any enhancement in the low energy 
spin fluctuations for the optimally doped systems
\cite{Ishidarev,Nakai,Fujiwara,Mukuda}. 
In the intermediate angle regime ($105<\alpha<115$), 
where the three hole Fermi surfaces coexist, 
superconductivity and the stoner factor is correlated, 
but in the smaller angle regime ($\alpha<105$), 
where $\alpha_1$ Fermi surface disappears, there 
is an complicated correlation between the two.
In this regime, the effect of the $\gamma$ Fermi surface
becomes even larger, overcoming the disappearance of the 
$\alpha_1$ Fermi surface and resulting in an increase of the stoner factor.
This works positively for superconductivity, and  
in fact, $\lambda$ is 
slightly enhanced in the angle regime $97<\alpha<100$ for $T=0.005$eV.
However, in the smaller bond angle regime, $a_S$ continues
to increase, but $\lambda$ is degraded. This is mainly because
all the Fermi surfaces, including the electron ones,  
become large in this regime.  
Namely, the relative size of the Fermi surface with respect to 
the spread of the spin fluctuations $\Delta Q$ becomes large, so that the pair
scattering processes occur only between restricted 
portions of the Fermi surfaces, while when the size of the Fermi
surfaces is comparable to  $\Delta Q$, 
the pair scattering can occur between the 
entire regime, as shown in
Fig.\ref{fig:3}(b)\cite{KurokiArita}. Thus, in the small angle regime, 
although the spin fluctuations at $(\pi,0)$ becomes stronger, 
superconductivity is degraded due to the disappearance of the 
$\alpha_1$ and the enlargement of other Fermi surfaces.

Finally we turn to the bond length dependence. 
In Fig.\ref{fig:2}(b), we plot the eigenvalue of the 
Eliashberg equation as a function of the bond length. 
The eigenvalue monotonically increases with the increase of the  
bond length\cite{comment}.
Since the change in the Fermi surface is small here, the enhancement of
the superconductivity is mainly due to the increase of the density of 
states originating from the narrowing of the band width.

In Fig.\ref{fig:2}(c), we present a schematic $T_c$ diagram 
in the bond angle and bond length space.
High $T_c$ is obtained when the 
bond angle is around 110 deg. (the regular tetrahedron angle) so that 
there are (nearly) three hole Fermi surfaces
and also when the bond length is 
large so that the band width is narrow\cite{KFeSe}.

To summarize, we have studied the condition for optimizing 
superconductivity in the iron pnictides from the lattice structure 
point of view. As found in Ca$_4$Al$_2$O$_6$Fe$_2$As$_2$\cite{Miyake}, 
the band structure of the hypothetical lattice structure of LaFeAsO also 
exhibits a disappearance of one of the hole Fermi surface in the  small 
bond angle regime, and therefore, the hole Fermi surface multiplicity is 
maximized around the bond angle regime where the arsenic atoms form a 
regular tetrahedron. Superconductivity is optimized within this 
three hole Fermi surface regime, while the $(\pi,0)$ stoner factor 
has an overall increasing tendency upon decreasing the bond angle.
Consequently, the correlation between superconductivity and the 
spin fluctuations becomes rather complicated.
Combining also the effect of varying the Fe-As bond length, 
we have provided a schematic $T_c$ diagram, which may give   
a guiding principle for obtaining higher $T_c$ materials.

We acknowledge Takashi Miyake and Kiyoyuki Terakura 
for motivating us to start the present study.
The numerical calculations were in part performed at the Supercomputer
Center, ISSP, University of Tokyo. This work was supported 
by Grants-in-Aid from MEXT, Japan. H.U. acknowledges support from JSPS.

\bibliography{perovskite}

\begin{thebibliography}{99}
\bibitem{kamihara_LaFeAsO}
Y.Kamihara, {\it et al.},
Am. Chem. Soc. {\bf 130}, 3296 (2008).
\bibitem{Ren}
Z-A Ren, {\it et al.}
, Phys. Lett. {\bf 25}, 2215 (2008).
\bibitem{Lee}
C.H. Lee {\it et al.},
, J. Phys. Soc. Jpn. {\bf 77}, 083704 (2008).
\bibitem{Kuroki_prb}
K. Kuroki, {\it et al.}
, Phys. Rev. B {\bf 79}, 224511 (2009)
\bibitem{XYZ} $X$, $Y$, $Z$ refer to the axes of the original unit cell 
containing two irons.
\bibitem{Mazin} I.I. Mazin {\it et al.}, 
Phys. Rev. Lett. {\bf 101}, (2008) 057003.
\bibitem{Graser} S.Graser {\it et al}, 
New J. Phys. {\bf 11}, 025016 (2009).
\bibitem{Kuroki_prl} K. Kuroki {\it et al.}, 
Phys. Rev. Lett. {\bf 101}, (2008) 087004, 
\bibitem{Miyazawa} K. Miyazawa {\it et al.}, J. Phys. Soc. Jpn. {\bf 78}, 
034712 (2009).
\bibitem{Takano} Y. Mizuguchi and Y. Takano, J. Phys. Soc. Jpn. {\bf 79}, 
102001 (2010) and references therein.
\bibitem{Ogino}
H. Ogino {\it et al., }
Supercond. Sci. Technol. {\bf 22}, 075008 (2009); 
{\it ibid} {\bf 23}, 115005 (2010).
\bibitem{Shirage}
P. M. Shirage {\it et al.}
,Appl. Phys. Lett. {\bf 97}, 172506 (2010)
\bibitem{Miyake}
T. Miyake {\it et al.}, J. Phys. Soc. Jpn. {\bf 79}, 123713 (2010)
\bibitem{Bickers} N.E. Bickers {\it et al.},
Phys. Rev. Lett. {\bf 62}, 961 (1989).
\bibitem{pwscf}
S. Baroni {\it et al.},
http://www.quantum-espresso.org/.
Here we adopt the exchange correlation functional introduced by
J. P. Perdew, 
(Phys. Rev. B {\bf 54}, 16533 (1996)), and the wave functions are expanded by
plane waves up to a cutoff energy of 40 Ry.
10$^3$ $k$-point meshes are used
with the special points technique by
H.J. Monkhorst and J.D. Pack (Phys. Rev. B {\bf 13}, 5188 (1976)).
\bibitem{MaxLoc} N. Marzari and D. Vanderbilt, Phys. Rev. B
{\bf 56}, 12847 (1997);
I. Souza, N. Marzari and D. Vanderbilt,
Phys. Rev. B {\bf 65}, 035109 (2001).
The Wannier functions are generated by the code developed by
A. A. Mostofi {\it et al},
(http://www.wannier.org/).
\bibitem{Singh} D.J. Singh and M.-H. Du: Phys. Rev. Lett. {\bf 100}, (2008) 
237003. 
\bibitem{Vildosola} V. Vildosola {\it et al.}, 
Phys. Rev. B {\bf 78}, 064518 (2008).
\bibitem{Miyake2}
T. Miyake {\it et al.}, J. Phys. Soc. Jpn. {\bf 79}, 044705 (2010)
\bibitem{Ikeda_jpsj}
H. Ikeda, J. Phys. Soc. Jpn. {\bf 77}, 123707 (2008)
\bibitem{Arita} R. Arita and H. Ikeda, J. Phys. Soc. Jpn. {\bf 78}, 113707 
(2009).
\bibitem{Ikeda_prb}
H. Ikeda {\it et al.}, Phys. Rev. B {\bf 81}, 054502 (2010)
\bibitem{DHLee} F. Wang {\it et al.}, 
Phys. Rev. B {\bf 81}, 184512 (2010).
\bibitem{Thomale} R. Thomale {\it et al.}, arXiv : 1002.3599.
\bibitem{Ishidarev} K. Ishida {\it et al.}, J. Phys. Soc. Jpn. {\bf 78}, 
062001 (2009) and references therein.
\bibitem{Nakai} Y. Nakai {\it et al.}, Phys. Rev. B {\bf 79}, 212506 (2009).
\bibitem{Mukuda} H. Mukuda {\it et al.}, J. Phys. Soc. Jpn. {\bf 78}, 
084717 (2009).
\bibitem{Fujiwara} T. Nakano {\it et al.}, Phys. Rev. B {\bf 81}, 
100510 (2010).
\bibitem{KurokiArita} 
K. Kuroki and R. Arita: Phys. Rev. B {\bf 64}, (2001) 024501.
\bibitem{comment} This monotonic enhancement of $\lambda$ 
is restricted to the regime of $f$ 
where the spin fluctuations develop at 
$(\pi,0)$. When large $f$ is taken, 
the spin fluctuation wave vector changes from $(\pi,0)$ to 
$(\pi,\pi)$ as the bond length is increased, and in that case 
$\lambda$ takes a maximum value at a certain bond length.
\bibitem{KFeSe} A recently discovered superconductor K$_x$Fe$_2$Se$_2$ 
(J. Guo {\it et al}, Phys. Rev. B {\bf 82}, 180520(R) (2010)) 
is also interesting from this viewpoint since the top of the 
three hole bands near $E_F$ are nearly degenerate and the band width is narrow.
\end{thebibliography}

\end{document}